\documentclass[]{spie}  

\usepackage{amsmath,amsfonts,amssymb}
\usepackage{graphicx}
\usepackage[colorlinks=true, allcolors=blue]{hyperref}

\title{Quick Ultra-VIolet Kilonova surveyor (QUVIK)}

\author[a]{N. Werner}
\author[a]{J. Řípa}
\author[a]{F. M\"unz}
\author[a]{F. Hroch}
\author[b]{M. Jelínek}
\author[a]{J. Krtička}
\author[a]{M. Zajaček}
\author[a,f]{M. Topinka}
\author[c]{V. Dániel}
\author[c]{J. Gromeš}
\author[d]{J. Václavík}
\author[d]{L. Steiger}
\author[d]{V. L\'edl}
\author[e]{J. Seginak}
\author[g]{J. Benáček}
\author[h]{J. Budaj}
\author[a]{N. Faltová}
\author[i]{R. Gális}
\author[a]{D. Jadlovský}
\author[a]{J. Janík}
\author[a]{M. Kajan}
\author[j]{V. Karas}
\author[k]{D. Korčáková}
\author[a]{M. Kosiba}
\author[a]{I. Krtičková}
\author[b]{J. Kubát}
\author[b]{B. Kubátová}
\author[a]{P. Kurf\"urst}
\author[a]{M. Labaj}
\author[a]{Z. Mikulášek}
\author[l]{A. Pál}
\author[a]{E. Paunzen}
\author[a]{M. Piecka}
\author[a,m]{M. Prišegen}
\author[a]{T. Ramezani}
\author[b]{M. Skarka}
\author[a]{G. Szász}
\author[b]{C. Th\"{o}ne}
\author[a]{M. Zejda}

\affil[a]{Department of Theoretical Physics and Astrophysics, Faculty of Science, Masaryk University, Brno, Czech Republic}
\affil[b]{Astronomical Institute (ASU CAS), Ond\v{r}ejov, Czech Republic}
\affil[c]{VZLU Czech Aerospace Research Centre, Prague, Czech Republic}
\affil[d]{Research Centre for Special Optics and Optoelectronic Systems (TOPTEC), Institute of Plasma Physics of the CAS, Prague, Czech Republic}
\affil[e]{PEKASAT SE, Brno, Czech Republic}
\affil[f]{INAF – Istituto di Astrofisica Spaziale e Fisica Cosmica, Milano, Italy}
\affil[g]{Center for Astronomy and Astrophysics, Technical University of Berlin, Berlin, Germany}
\affil[h]{Astronomical Institute, Slovak Academy of Sciences, Tatranská Lomnica, Slovak Republic}
\affil[i]{Institute of Physics, Faculty of Science, Pavol Jozef Šafárik University, Košice, Slovakia}
\affil[j]{Astronomical Institute of the Czech Academy of Sciences, Prague, Czech Republic}
\affil[k]{Astronomical Institute, Faculty of Mathematics and Physics, Charles University, Praha, Czech Republic}
\affil[l]{Konkoly Observatory, Research Centre for Astronomy and Earth Sciences, Eötvös Loránd Research Network (ELKH), Budapest, Hungary}
\affil[m]{Advanced Technologies Research Institute, Faculty of Materials Science and Technology in Trnava, Slovak University of Technology in Bratislava, Slovakia}

\authorinfo{Further author information: (Send correspondence to NW)\\NW: E-mail: werner@physics.muni.cz}

\begin{document} 
\maketitle

\begin{abstract}
We present a near-UV space telescope on a $\sim$70kg micro-satellite with a moderately fast repointing capability and a near real-time alert communication system that has been proposed in response to a call for an ambitious Czech national mission. The mission, which has recently been approved for Phase 0, A, and B1 study shall measure the brightness evolution of kilonovae, resulting from mergers of neutron stars in the near-UV band and thus it shall distinguish between different explosion scenarios. Between the observations of transient sources, the satellite shall perform observations of other targets of interest, a large part of which will be chosen in open competition. 
\end{abstract}

\keywords{kilonovae, ultraviolet, space telescope, micro-satellite}

\section{INTRODUCTION}
\label{sec:intro}  

All elements heavier than lithium were moulded in stars. Elements like C, N, O, Si, Fe, and Ni were produced in stellar furnaces and spread across the Universe by stellar outflows and supernova explosions. However, the even heavier and much rarer elements, like gold and platinum, must have been created by a different process, for example, when neutron-rich matter released from an explosive event underwent rapid neutron capture nucleosynthesis, the so-called r-process \cite{burns2020}. The r-process appears to be particularly efficient during the coalescence of double neutron stars (NS-NS), which are also prime sources of gravitational waves (GW) for detectors like the Advanced LIGO, Virgo, KAGRA and others \cite{abbott2020}. Such events result in so-called kilonovae \cite{metzger2019}, powered by the radioactive decay of unstable heavy nuclei. 

The current state-of-the-art in observational studies of kilonovae and r-process nucleosynthesis are the observations of the electromagnetic counterpart of the NS-NS merger, GW170817 \cite{abbott2017}, discovered by LIGO/Virgo \cite{margutti2021}. These observations revealed a blue transient, with the emission peaking at near-UV wavelengths, and the subsequent optical and infrared observation showed a redward evolution of the emission over about 10 days. These observations indicate that the merger indeed resulted in a kilonova, which might have produced as much as 0.05 Solar masses of ejecta. Kilonovae thus might be one of the prime sites for the production of elements heavier than Ni in the Universe. However, the optical counterpart of GW170817 was only discovered by telescopes 11 hours after the coalescence \cite{abbott2017} and earlier observations, which are necessary for our understanding of the physics of kilonovae, are unavailable. A wide-field near-UV space telescope with a fast-repointing capability, enabling observations of the early emission, would result in real breakthroughs for our understanding of the nucleosynthesis of kilonovae.

The most powerful current UV space observatory is the Hubble Space Telescope. Other UV observatories include the Ultraviolet/Optical Telescope (UVOT) \cite{roming2005} on the Neil Gehrels Swift Observatory and the Ultra Violet Imaging Telescope (UVIT) on the Indian AstroSat mission \cite{agrawal2017}. All these observatories are well beyond their planned lifetimes and currently there are no approved UV observatories in Europe, US, or Japan. In early 2025, Israel is planning to launch the ULTRASAT \cite{sagiv2014} mini-satellite carrying a telescope with 33 cm aperture and a very large field of view of 200 square degrees. It is optimised for a single ultraviolet band of 220-280 nm and is planned to operate on geostationary orbit (GEO). ULTRASAT is expected to reach a 5$\sigma$ sensitivity of 22.3 AB limiting magnitude in 3x300s integrations. 

To study the physics of kilonovae, we proposed a mission that is complementary to ULTRASAT. The Quick Ultra-VIolet Kilonova surveyor (QUVIK) is a small but agile and capable mission carrying a sensitive telescope to provide near-UV photometry of kilonovae in two bands early after their explosion. The project is lead by the Czech Aerospace Research Centre VZLU, which is also responsible for the platform design. The Department of Theoretical Physics and Astrophysics in the Faculty of Science at Masaryk University leads the scientific preparations, including the definition of the scientific objectives and requirements. The Research Centre for Special Optics and Optoelectronic Systems TOPTEC is responsible for the telescope design. PEKASAT is responsible for the design of the communication solutions, ground station, and mission operation centre.

\section{Early photometry of kilonovae as the main science driver }

Early near-ultraviolet observations of the electromagnetic counterpart of GW170817 revealed a blue transient and the subsequent optical and infrared observations showed a redward evolution of the emission over about 10 days. These observations provide a strong support to the hypothesis that the gravitational wave signal GW170817 was produced by the merger of two neutron stars, resulting in a kilonova powered by r-process nuclei synthesised in the ejecta. As we already mentioned in the introduction, the observation also provided important indications that neutron star mergers are the primary sources of rare heavy elements produced by the r-process \cite{metzger2019}. These results are of key importance in astrophysics and no similar observations were performed since August 2017.

The estimated rates of mergers involving neutron stars within the distance of 200 Mpc range from a few to a few tens per year \cite{abbott2020}. With further upgrades the sensitivity of LIGO, Virgo, and KAGRA will continue to increase. By the mid-2020s, it is expected that advanced LIGO will be able to detect NS-NS mergers up to a distance of 330 Mpc, advanced Virgo to 150-260 Mpc, and KAGRA to a distance larger than 130 Mpc \cite{abbott2020}. Importantly, the localisation accuracy will be between 5 deg$^2$ and 20 deg$^2$ \cite{abbott2018}. Therefore, after receiving an alert, the satellite will start performing mosaic observations of the potential target area until a more precise localisation is available. For successful mosaicking and follow-up observations, a large field of view of at least $1.5\times1.5$ deg$^2$ is required.

While GW170817 was quickly followed by a short GRB seen at an angle of 19-42 degrees from the jet axis, it is likely that most kilonovae will not emit a GRB observable from Earth \cite{metzger2019}. The prompt gamma-ray emission is strongly directional, and it is estimated that only about 1 in 100 kilonovae will be detectable at high energies \cite{metzger2012}. However, the detection of another short GRB counterpart for a gravitational wave event would be an important discovery. We will therefore consider equipping our proposed UV satellite with a hosted GRB detector, that is also capable to perform degree-scale localisation. Such a detector could enable us to perform a quick follow-up of those rare kilonovae, which are associated with an observable GRB. 
 
Rapid follow-up observations at near-UV wavelengths are expected to produce real breakthroughs in our understanding of kilonovae.  
A NS-NS merger can lead to the formation of a hyper-massive neutron star, which quickly collapses to a black hole; to a stable, rapidly spinning, highly magnetised neutron star; or directly to a black hole. The merger also results in $10^{-4}$--$10^{-2}$ Solar masses of ejecta \cite{metzger2012}. The equatorial ejecta is expected to be rich in heavier elements, known as lantanides, and produce long-lasting infrared emission. On the other hand, free neutrons and lantanide poor polar ejecta, with a higher electron fraction, result in UV and blue emission that last for only about a day. The flux ratios observed in our proposed near-UV bands, and at optical and near infrared wavelengths observable from the ground in the first hours/day of the kilonova, will allow us to probe the mass, composition, and thermal content of the fastest ejecta and determine the nature of the merger product \cite{metzger2019}.

A micro-satellite carrying a near-UV space telescope with roughly 200 cm$^2$ effective arrea (after correcting for losses in the optical system and detector quantum efficiency) will be able to provide a S/N=5 detection of a 22 mag object using a 12 min total exposure (possibly coadded from much shorter exposures) and thus will probe the emission from kilonovae out to the distance of up to 200 Mpc. The moderately fast repointing capability of the small and light micro-satellite will enable us to follow up the target within a few tens of minutes. This capability will lead to unprecedented data about the early emission of kilonovae.

\section{Important science between kilonova observations}

The satellite will also perform fast target-of-opportunity follow-up observations of GRB afterglows. These observations will allow us to improve the localisation of GRBs, as well as identify and determine their precise position within their host galaxy. This information will help perform ground-based follow-up spectroscopic measurements to determine the host galaxy’s redshift. Importantly, two-band UV photometry of early GRB afterglows will help us test various models on the physics of GRB jets. 

Next to kilonovae and GRBs, the satellite will also perform target of opportunity observations of supernovae. Early UV observations of type Ia supernovae, performed less than 1 day after the explosion, will be particularly important for determining whether the progenitor is single-degenerate (white dwarf accreting from a non-degenerate companion) or double-degenerate (coalescence of two white dwarfs). Single-degenerate explosion scenarios predict a UV flare when the supernova ejecta interacts with the companion star \cite{cao2015}. Next to supernovae, target of opportunity observations of cataclysmic variables will also present important secondary science. 

Target-of-opportunity observations of tidal disruption events (TDEs), which arise when a star gets disrupted by the central black hole of a galaxy, is another very promising area for first-class secondary science. It turns out that TDEs are relatively frequent: in a Milky-Way like galaxy, they occur about once per $\gtrsim10^4$ years. For the typical number density of Milky Way-like galaxies of 0.006 Mpc$^{-3}$, this results in $\lesssim24$ TDEs per year up to the redshift of 0.05. This is in agreement with $\sim14$ TDEs per year detected by the Zwicky Transient Facility optical survey \cite{Bellm2019}. Surveys in the coming decade (including ULTRASAT with its large field of view) are expected to discover many TDEs providing triggers for follow-up observations. Two-band UV follow-up observations may provide confirmation of TDEs, telling them apart from a weak AGN activity \cite{vanvelzen2020}.

While the satellite will be standing by and not pointing at transients, it could observe and monitor other targets, such as transiting exoplanets, stars or AGNs. UV observations of exoplanet transits are expected to be particularly important for the understanding of their atmospheres. Observations with e.g., SWIFT/UVOT indicate significant differences in the inferred exoplanet sizes determined using optical and NUV observations \cite{salz2019}. Deep NUV transits appear to be positively correlated with mass-loss rates of planetary atmospheres \cite{owen2019}. The UV variability of stars remains largely unexplored, therefore a survey of hot stars and stellar populations in star clusters is expected to provide important new results. Surveys of galaxies, which will allow us to place constraints on star-formation rates and study galaxy evolution, are also among promising research topics. Two-band quasi-simultaneous UV observations, supplemented by concurrent optical data, of nearby AGN with a high cadence of $<0.5$ days can be utilized for photometric reverberation mapping of accretion discs. This will be critical for probing known accretion-disc models as well as for determining accretion-disc sizes \cite{Kumar2022}.
 
In order to maximise the scientific utilisation of the proposed UV telescope, we will offer open time for scientists from all around the world. The satellite will thus be used for the best proposed science by a world-wide community. Furthermore, after a 1 year long proprietary time, we will make all the collected data freely available to anyone interested in the scientific exploitation of the performed observations. 

\section{Mission design}

   \begin{figure} [ht]
   \begin{tabular}{c} 
   \includegraphics[height=6cm]{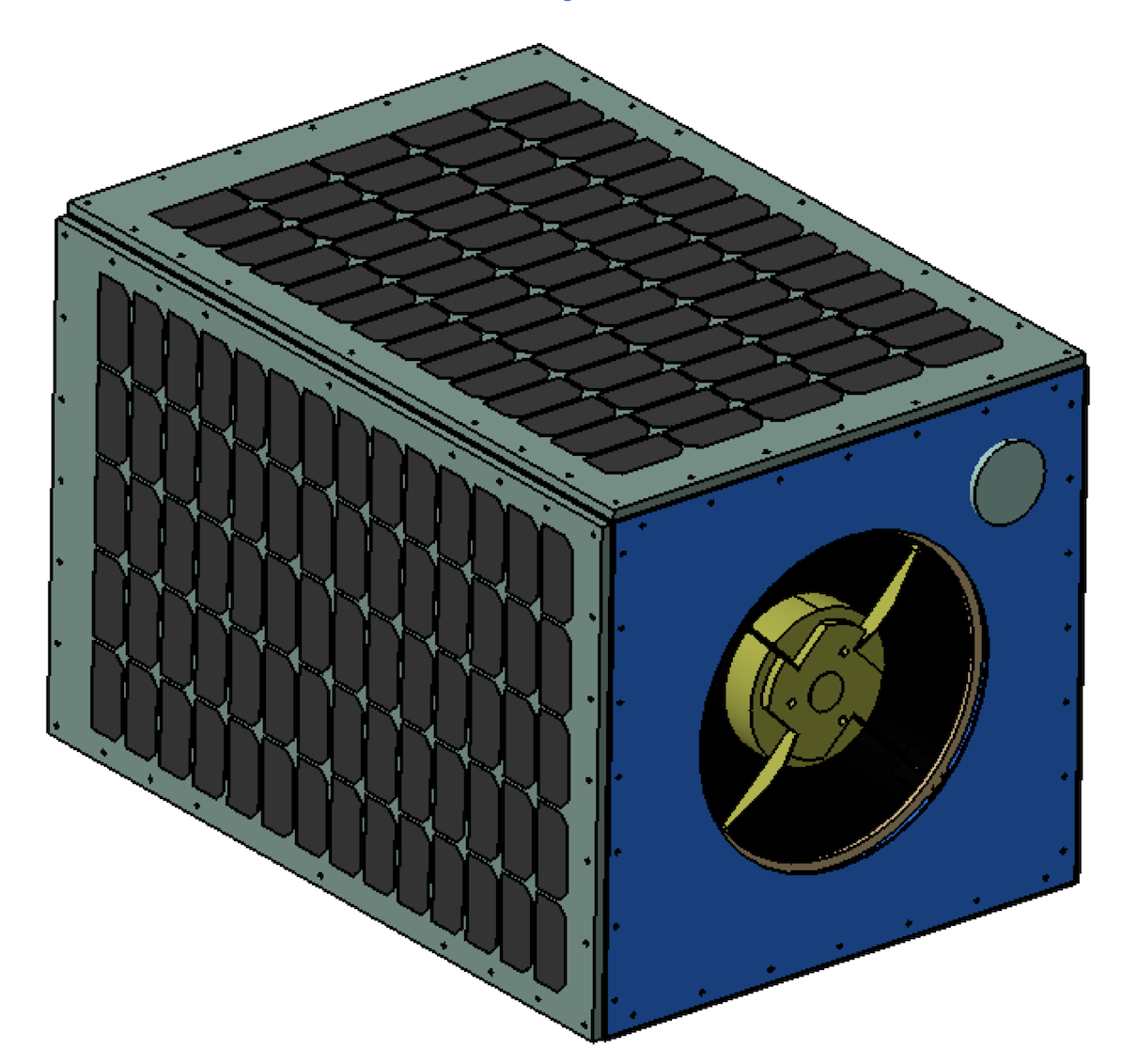}
   \hspace*{3cm}   \includegraphics[height=6.3cm]{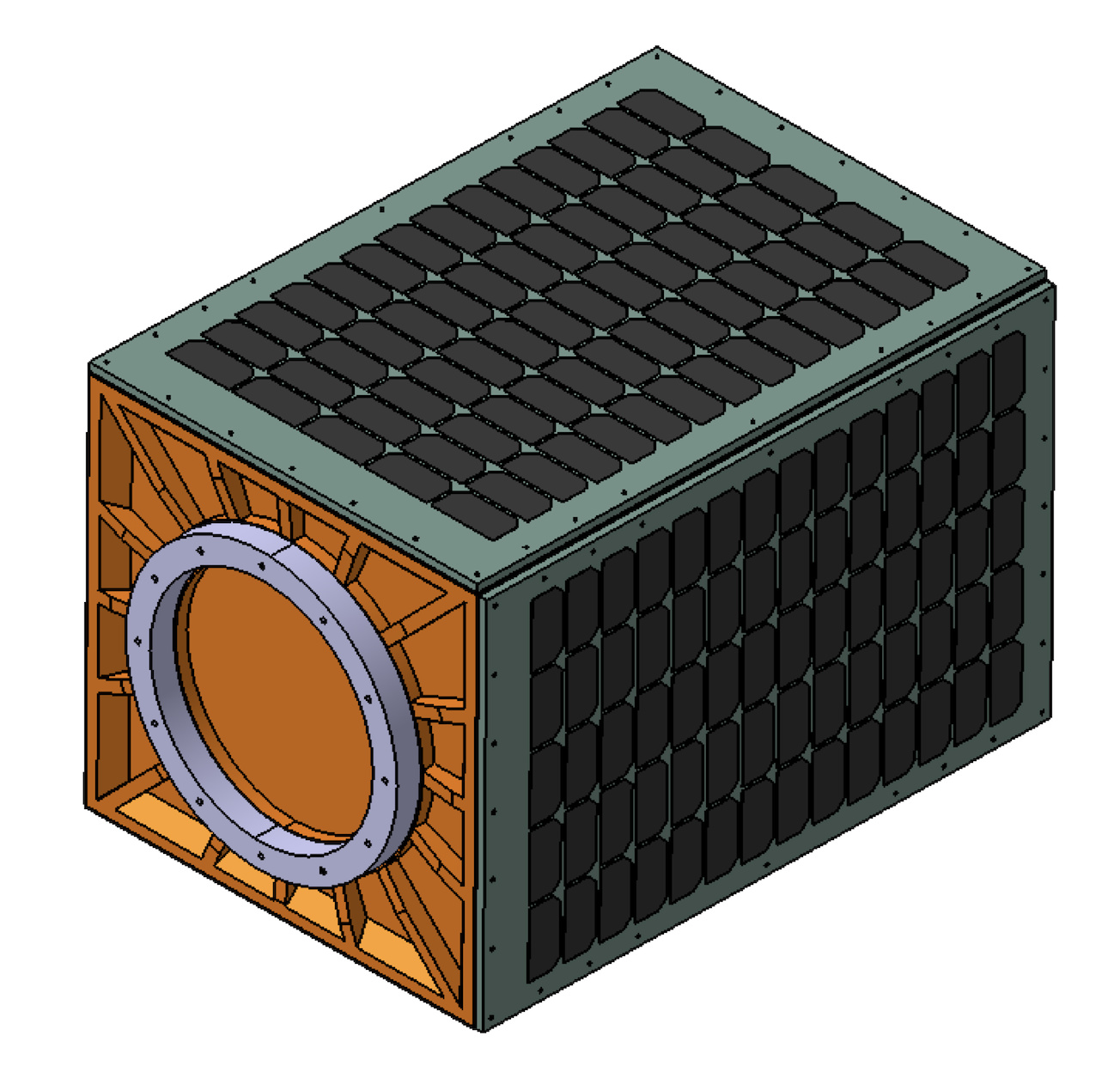}
	\end{tabular}
   \caption[example] 
   {An illustration of the QUVIK micro-satellite, hosting a near-UV space telescope on-board, from two different directions.}
   \end{figure}

Following a detection of a gravitational wave or GRB event by existing or future instruments, an alert will be issued in near-real time. The alert will be processed by the Mission Operation Centre (MOC) and if observable (unobstructed by the Earth or the Sun), the coordinates of the event will be transferred as an alert via a 3rd party constellation to the spacecraft. Subsequently, the spacecraft will repoint to start the observation of the source of the event. This sequence shall be performed within tens of minutes from the time when the alert is issued. 

The micro-satellite (see Fig. 1) on a low-Earth orbit will be equipped with a telescope with a relatively large field of view, with a diameter of at least 2 degrees, to facilitate the localisation of transients. We expect that in the last years of this decade, many gravitational wave sources, that involve a neutron star, will be localised with a sufficient accuracy to allow the error region containing the transient to be imaged with a relatively few pointings. The imaging will be performed in two near-UV bands. 

Since most relatively affordable launch opportunities, such as the Transporter ride-share program by SpaceX, tend to aim for the  sun-synchronous polar orbit (SSO), this is also the most likely orbit of QUVIK. 

\subsection{The on-board telescope}

The payload is based on an optical telescope assembly (OTA) and one or multiple imaging units. The imaging unit controls the imaging sensor, buffers acquired data and provides asynchronous communication over the satellite bus through the payload controller. The core of the imaging unit is an image sensor. A set of commercial off-the-shelf CCDs (Teledyne e2v, Hamamatsu) and CMOS (GSense) detectors with the required near-UV sensitivity has been identified. Sensors with 4096×4096 resolution were chosen as a baseline for further evaluation due to angular resolution and field-of-view requirements.

If multiple exposures are needed due to the attitude stability limitations, the CMOS sensor will be preferred. However, the high speed, lower readout noise and smaller pixel size of the CMOS sensors are offset by lower sensitivity, only 12-bit ADC and lower flexibility compared to the CCDs. 

The OTA design parameters are based on the selected image detector. The OTA contains a dichroic beam splitter and additional filters forming two images in two spectral bands. The optical design of the telescope is driven by the requirements on the effective area, field of view, and spectral range. A modified Cassegrain design (hyperbolic primary, aspheric secondary mirrors) with a focal plane corrector was selected as a baseline to evaluate the optical and technical requirements (see Fig. 2). The lenses in the correction optics are made of fused silica, MgF$_2$ and CaF$_2$, which are readily available. Our conceptual design offers better than required performance in a nominal state (e.g. without manufacturing errors, misalignments, or thermal effects). Detailed analysis will be performed to find the optimal design from the point of view of optical, mechanical, thermal performance and manufacturing constraints.

   \begin{figure} [ht]
   \begin{center}
   \begin{tabular}{c} 
   \includegraphics[height=8cm]{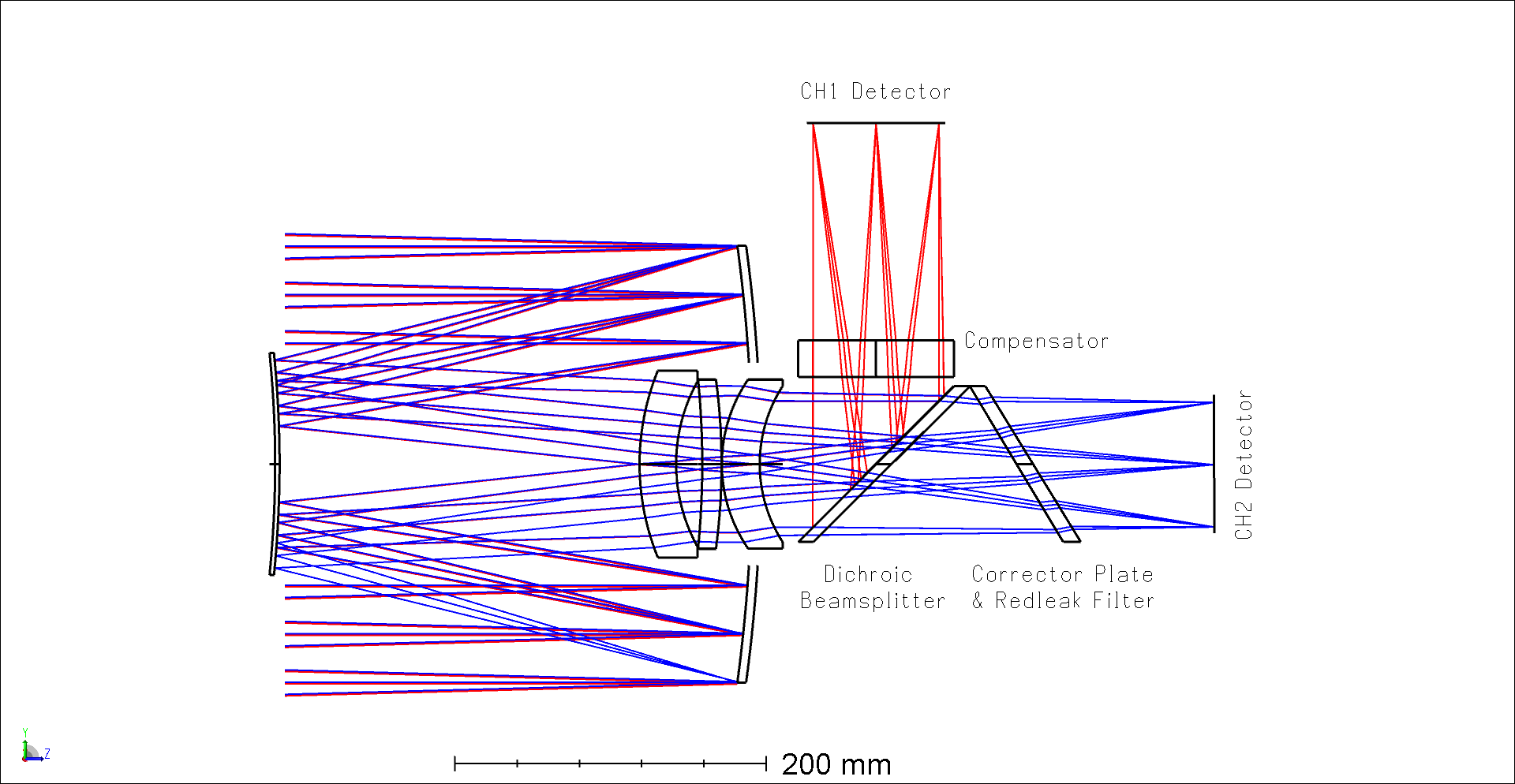}
	\end{tabular}
	\end{center}
   \caption[example] 
   {Preliminary design of a modified dual-band telescope based on a modified Cassegrain design with a common 3-lens field corrector and beam splitter assembly. The colors represent the different spectral bands. }
   \end{figure} 

Lenses are designed with spherical surfaces to allow the use of full-aperture polishing technology on difficult to polish materials (e.g. MgF$_2$, CaF$_2$). The OTA will require special, high-efficiency coatings of optical surfaces to reach the required efficiency. The performance estimation is based on commercially available coatings. The estimated efficiency of the OTA is 41\%, resulting in an effective collection area of 200 cm$^2$. The mechanical construction of the baseline design is based on a truss structure. The primary mirror is firmly connected to an optical bench containing corrective optics, a beamsplitter and the focal plane assembly. The secondary mirror is held by the truss structure. The telescope will be athermalized by the combination of a truss material (e.g. SiC + aluminium compensator) and material of the corrector barrel and optical bench (e.g. titanium). Alternative athermalization method is to use a single material for the truss structure (e.g. SiC) and a different material for the primary and the secondary mirror (e.g. Pyrex). Optical baffles and wanes will be formed from a thin metal sheet attached to the truss structure.

\subsection{Communication}

The nominal telemetry and commanding shall be performed via S-band radio. A UHF radio is being baselined as a backup for telemetry. S-band radio offers higher data rates than UHF and can be used to perform on-board software updates. The UHF radio channel places lower requirements on pointing accuracy than S-band and it can thus be useful for the time period following spacecraft separation. It can also be used in case of an attitude control system malfunction.

For downloading science data, a high-speed downlink is required. Our preliminary studies indicate that an X-band radio will provide an optimal solution. 

Sending the coordinates of transient events to the spacecraft and thus enabling their quickest possible follow-up observations will require a near real-time uplink \& downlink possibility. For that, we expect to use a 3rd party inter-satellite network. At lower latitudes, we could use solutions based on GEO constellations, such as the Intersatellite Data Relay System (ISDR). At higher latitudes, we will study the use of Iridium NEXT, Orbcom, and Globalstar networks.

\section{Summary}

The Quick Ultra-VIolet Kilonova surveyor (QUVIK) mission has been approved for Phase 0, A, and B1 study, which shall be completed by June 2023. If approved for implementation, the mission shall be ready for launch around 2026/2027. While the primary science objective of the satellite, which drives its design development, is the early photometry of kilonovae, it is expected that its observing program will be dominated by other interesting science, such as the study of gamma-ray burst afterglows, supernovae, hot stars, binaries, exoplanet transits, star clusters, as well as galaxies and their nuclei. A large portion of this observing time will be open to the worldwide science community, which will be able to submit proposals for observations. After a one year-long proprietary time, all the collected data will be made freely available online.

\bibliography{report} 
\bibliographystyle{spiebib} 

\end{document}